\documentstyle[floats,aps,epsf,prl]{revtex}

\begin{document}
\draft

\twocolumn[\hsize\textwidth\columnwidth\hsize\csname
@twocolumnfalse\endcsname

\title{Berry-phase theory of proper piezoelectric response}
\author{David Vanderbilt}
\address{Department of Physics and Astronomy, Rutgers University,
Piscataway, New Jersey 08855-0849}

\date{March 1, 1999}
\maketitle

\begin{abstract}
Recent theoretical advances have established that the electric
polarization in an insulating crystal can be viewed as a
multivalued quantity that is determined by certain Berry phases
associated with the occupied Bloch bands.  The application of this
approach to the computation of piezoelectric coefficients is
not entirely straightforward, since a naive determination
of the (``improper'') piezoelectric coefficients from
finite differences of the polarization at nearby strain states
leads to a dependence upon the choice of ``branch'' of the
polarization.  The purpose of the present paper is to clarify
that if one calculates instead the ``proper'' piezoelectric
response, the branch dependence is eliminated.  From this
analysis, a simplified recipe for the direct finite-difference
computation of the proper piezoelectric coefficients emerges
naturally.
\end{abstract}

\vskip1pc]

\narrowtext


\section{Introduction}
\label{sec:intro}

The calculation of spontaneous polarization and piezoelectric response
within the framework of first-principles methods of electronic
structure theory has proven to be a rather subtle problem.
In a landmark paper, Martin \cite{martin} showed that the
piezoelectric tensor {\it is} well-defined as a bulk quantity in
a crystalline insulator.  However, at that time it was far from
clear whether the spontaneous polarization itself could be regarded
as a bulk property in the same sense, and calculations of piezoelectric
constants by finite differences of spontaneous polarization were
therefore not possible.

The situation changed in 1993 with the development of the
``Berry-phase'' theory of polarization \cite{ksv,vks}, which provided
a direct and straightforward method for computing the electric
polarization.  (For a useful review, see Ref.~\onlinecite{resta}.)
Nevertheless, some subtleties remain regarding the computation
of the piezoelectric tensor components by finite differences
\cite{sck,bernardini}.  First, the Berry-phase theory gives the
polarization as a multivalued quantity, and the piezoelectric
response that would be computed from a given one of the many
branches is not invariant with respect to choice of branch.
Second, a distinction is made between the ``proper'' and
``improper'' piezoelectric response \cite{martin,nl,nelson},
and it might not be clear which of these is to be associated with
the finite-difference calculation.

The purpose of the present paper is to elucidate the physics of
the spontaneous polarization, the piezoelectric response, and the
relations between the two.  It is clarified that the improper
polarization is the one given by the naive finite-difference
approach, and that while this quantity is indeed branch-dependent,
the proper polarization, which should be compared with experiment,
is not.  As a result of this analysis, a simplified recipe for the
direct finite-difference computation of the proper piezoelectric
response is given.


\section{Berry-phase theory of polarization}
\label{sec:berry}

We consider a periodic insulating crystal in zero macroscopic
electric field, and assume that the electronic ground state can
be described by a one-electron Hamiltonian $H$ as in density-functional
or Hartree-Fock theory.  The eigenstates of $H$ are the Bloch
functions $\psi_{n\bf k}$ with energies $\epsilon_{n\bf k}$,
and it is conventional to define the cell-periodic Bloch functions
\begin{equation}
u_{n\bf k}({\bf r})=
e^{-i\bf k\cdot r}\,\psi_{n\bf k}({\bf r})
\label{eq:u}
\end{equation}
having periodicity $u_{n\bf k}({\bf r})=u_{n\bf k}({\bf R+r})$,
where $\bf R$ is any lattice vector.  The contribution of the
$n$'th occupied band to the spontaneous electric polarization
of the crystal can then be written \cite{ksv,vks}
\begin{equation}
{\bf P}_n = {ie\over(2\pi)^3}\,\int d^3k\,
  \langle u_{n\bf k}\vert\nabla_{\bf k}\vert u_{n\bf k}\rangle
  \;\;.
\label{eq:P}
\end{equation}
We take the convention that $n$ runs over bands and spin, so a
factor of two would need to be inserted in Eq.~(\ref{eq:P})
to account for paired spins.  The total spontaneous polarization
is then given by
\begin{equation}
{\bf P} = {e\over\Omega}\,\sum_\tau\,Z_\tau\,{\bf r}_\tau
  \;+\;\sum_{n\;{\rm occ}}\,{\bf P}_n
  \;\;,
\label{eq:Ptot}
\end{equation}
where $Z_\tau$ and ${\bf r}_\tau$ are the atomic number and
cell position of the $\tau$'th nucleus in the unit cell, and
$\Omega$ is the unit cell volume.

Strictly speaking, Eq.~(\ref{eq:P}) applies only to an isolated
band, i.e., a band for which $\epsilon_{n\bf k}$ does not become
degenerate with any other band at any point in the Brillouin
zone.  This restriction is not essential; methods for extending
the analysis to composite groups of occupied bands containing
arbitrary degeneracies and crossings have been developed as
described in Refs.~\onlinecite{ksv,vks,resta}.  However, for
simplicity of presentation, it will be assumed here that only
isolated bands are present.  For the same reason, spin degeneracy
is suppressed throughout.

There is a certain arbitrariness inherent in Eq.~(\ref{eq:P})
associated with the freedom to choose the phases of the Bloch
functions $\psi_{n\bf k}$.  For, suppose we make a different choice
\begin{equation}
\vert\widetilde{\psi}_{n\bf k}\rangle=e^{i\beta({\bf k})}
\vert\psi_{n\bf k}\rangle
\;\;.
\label{eq:gauge}
\end{equation}
We shall refer to this as a ``gauge transformation'' of the Bloch
functions.  Note that the choice of $\beta{(\bf k})$ is restricted
by the fact that $\bf k$ and $\bf k+G$ label the same wavefunction
(where $\bf G$ is a reciprocal lattice vector), so that $\beta({\bf
k+G})-\beta({\bf k})$ must be an integer multiple of $2\pi$ for
any $\bf G$.  Thus, the most general form of $\beta(\bf k)$ is
\begin{equation}
\beta({\bf k})=\beta_{\,\rm per}({\bf k})+\bf k\cdot R
\label{eq:beta}
\end{equation}
where $\beta_{\,\rm per}$ is a periodic function in $k$-space and
$\bf R$ is some real-space lattice vector.  Letting $\widetilde{\bf P}_n$
be the result of inserting the $\widetilde{u}_{n\bf k}$ in place
of the $u_{n\bf k}$ in Eq.~(\ref{eq:P}), and using Eqs.~(\ref{eq:u}),
(\ref{eq:gauge}), and (\ref{eq:beta}), one finds
\begin{equation}
\widetilde{\bf P}_n={\bf P}_n-{e{\bf R}\over\Omega}
\;\;.
\label{eq:tildeP}
\end{equation}
Thus, while the contribution of this band to the electronic
polarization is not absolutely gauge-invariant, it is
gauge-invariant modulo $e/\Omega$ times a real-space lattice
vector.  Actually, this is precisely the type of qualified
invariance we should have expected.  After all, the choice of
the location $\bf r_\tau$ of the atom representing sublattice
$\tau$ in the unit cell has a similar ambiguity; we could just
as well choose $\widetilde{\bf r}_\tau={\bf r}_\tau+\bf R'$,
where $\bf R'$ is another lattice vector, leading to precisely
the same kind of ``modulo $e{\bf R}/\Omega$'' ambiguity
in the expression for $\bf P$ in Eq.~(\ref{eq:Ptot}).

Perhaps the most natural way to incorporate this kind of
ambiguity in the definition of the polarization is to regard
$\bf P$ as a {\it multivalued} quantity; that is, it simultaneously
takes on a {\it lattice} of values given by some ${\bf P}^{(b)}$
(here `$b$' is a ``branch'' label)
and all its periodic images ${\bf P}^{(b)}+e{\bf R}/\Omega$
(with $R$ running over all lattice vectors of the crystal).
To interpret this intuitively, we can say that from the point
of view of its dipolar properties, the real insulator behaves like
a fictitious crystal composed of two sublattices of point
$\pm e$ charges, with the sublattice of $-e$ charges displaced
relative to the $+e$ sublattice by $-\Omega{\bf P}/e$.  That is,
choosing one of the $+e$ charges as the origin, $-\Omega{\bf P}/e$
takes on a lattice of values that is precisely the lattice of
positions of the $-e$ charges.

\begin{figure}
\epsfxsize=3.4in
\centerline{\epsfbox{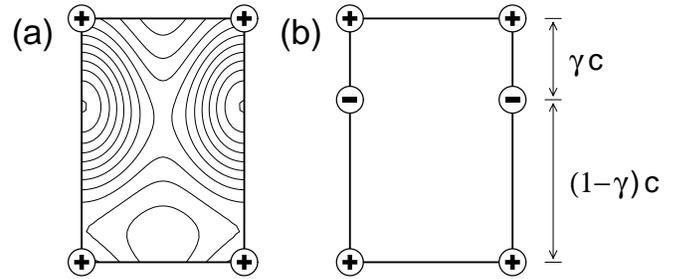}}
\vskip 0.5cm
\caption{Illustrative tetragonal crystal (cell dimensions
$a\times a\times c$) having one monovalent ion at the cell corner
(origin) and one occupied valence band.  (a) The distributed
quantum-mechanical charge distribution associated with the
electron band, represented as a contour plot.  (b) The
distributed electron distribution has been replaced by a unit
point charge $-e$ located at the Wannier center ${\bf r}_n$, as
given by the Berry-phase theory.}
\label{fig1}
\end{figure}

This is illustrated in Fig.~1
for an imaginary tetragonal crystal (dimensions $a\times a\times c$)
with one monovalent ion located at the cell corners, and a
single (spinless) electron band giving rise to the distributed
electron charge indicated schematically by the contours in
Fig.~1(a).  (We assume that $M_z$ mirror symmetry is broken in
some way.) Eq.~(\ref{eq:P}) then gives the location
\begin{equation}
{\bf r}_n=-\Omega{\bf P}_n/e
\label{eq:rn}
\end{equation}
of the effective unit point charge $-e$ illustrated in Fig.~1(b).
As discussed in Refs.~\onlinecite{ksv,vks,resta}, this location
is just the charge center of the Wannier function associated with
the electron band.  The polarization will then take on a lattice
of values having $x$, $y$, and $z$ components of $m_1e/ac$,
$m_2e/ac$, and $(\gamma+m_3)e/a^2$, respectively, where
the $m_i$ are integers.  More generally, when several occupied
bands are present, one can rewrite Eq.~(\ref{eq:Ptot}) as
\begin{equation}
{\bf P} = {e\over\Omega}\,\sum_\tau\,Z_\tau\,{\bf r}_\tau
  \;-\;{e\over\Omega}\,\sum_{n\;{\rm occ}}\,{\bf r}_n
  \;\;.
\label{eq:Ptotnew}
\end{equation}
%


In practice, one proceeds by computing the component of $\bf P_n$
along a particular crystallographic direction $\alpha$ via the
quantity
\begin{equation}
\phi_{n,\alpha}=-{\Omega\over e}{\bf G}_\alpha\cdot{\bf P}_n
\;\;,
\label{eq:phin}
\end{equation}
where $\bf G_\alpha$ is the primitive reciprocal lattice vector
in direction $\alpha$.
In cases of simple symmetry (e.g., tetragonal or rhombohedral
ferroelectric phases), a single $\phi_n$ suffices to
determine ${\bf P}_n$, but in general ${\bf P}_n$ can be
reconstructed from the three $\phi$'s via
\begin{equation}
{\bf P}_n=-{1\over2\pi}{e\over\Omega}\sum_\alpha \phi_{n,\alpha}{\bf R}_\alpha
\;\;,
\label{eq:buildP}
\end{equation}
where ${\bf R}_\alpha$ is the real-space primitive lattice vector
corresponding to ${\bf G}_\alpha$.  The $\phi_{n,\alpha}$ are angle
variables (``Berry phases'') that are well-defined modulo $2\pi$,
given by
\begin{equation}
\phi_{n,\alpha}=\Omega_{\rm BZ}^{-1}\int_{\rm BZ} d^3k\,
\langle u_{n\bf k}\vert
   -i{\bf G}_\alpha\cdot\nabla_{\bf k}
\vert u_{n\bf k}\rangle
\;\;,
\label{eq:phidef}
\end{equation}
where $\Omega_{\rm BZ}=(2\pi)^3/\Omega$ is the Brillouin zone (BZ)
volume.

The $\phi_{n,\alpha}$ can be regarded as giving the position of the
Wannier center for band $n$.  For the toy crystal of Fig.~1, for
example, and with the origin chosen at the cell corner, one would
have $\phi_x=\phi_y=0$ and $\phi_z=-2\pi\gamma$.
The practical calculation of the $\phi_{n,\alpha}$
proceeds on a discrete mesh in reciprocal space, arranged as a
two-dimensional grid of ${\bf G}_\alpha$-oriented strings
of k-points, as described in Refs.~\onlinecite{ksv,vks,resta}.


\section{Piezoelectric response}
\label{sec:piezo}

The piezoelectric tensor of a crystal reflects the first-order
change in spontaneous electric polarization in response to a
first-order deformation of the crystal.  The ``improper''
piezoelectric tensor is defined as \cite{nl,nelson}
\begin{equation}
c_{ijk}={dP_i\over d\epsilon_{jk}}
\label{eq:improper}
\end{equation}
in terms of the deformation
\begin{equation}
dr_j=\sum_k \,d\epsilon_{jk} \, r_k
\;\;,
\label{eq:deform}
\end{equation}
where the symmetric and antisymmetric parts of $d\epsilon$ represent
infinitesimal strains and rotations, respectively.  On the
other hand, the ``proper'' piezoelectric tensor can be defined as
\begin{equation}
\widetilde{c}_{ijk}={dJ_i\over d\dot{\epsilon}_{jk}}
\;\;,
\label{eq:proper}
\end{equation}
where $\bf J$ is the current density that flows through the bulk of
the sample in adiabatic response to a slow deformation
$\dot{\epsilon}=d\epsilon/dt$.  According to the standard references
\cite{nl,nelson}, the relation between the improper and proper
piezoelectric tensors is
\begin{equation}
\widetilde{c}_{ijk}=c_{ijk} + \delta_{jk}P_i - \delta_{ij} P_k
\;\;.
\label{eq:relation}
\end{equation}
Writing out explicit tensor components, this last equation becomes
\cite{sck}
\begin{eqnarray}
\widetilde{c}_{zzz} &=& c_{zzz} \;\;,\nonumber\\
\widetilde{c}_{zxx} &=& c_{zxx} \,+\,P_z \;\;,\nonumber\\
\widetilde{c}_{zxy} &=& c_{zxy} \;\;,\nonumber\\
\widetilde{c}_{zxz} &=& c_{zxz} \;\;,\nonumber\\
\widetilde{c}_{zzx} &=& c_{zzx} \,-\,P_x \;\;,
\label{eq:writeout}
\end{eqnarray}
and similarly for permutations of the cartesian labels (but not
for permutations of their position in the index triplet).
It might seem strange at first sight that the expressions for
$\widetilde{c}_{zxz}$ and $\widetilde{c}_{zzx}$ have a different
form, but this just reflects the fact that the deformation tensor
$\epsilon$ has been allowed to contain an antisymmetric part.

Now in the Berry-phase theory, the polarization is a multivalued
quantity, so that any particular value ${\bf P}^{(b)}$ has to
be identified by its branch label `$b$', and the corresponding
improper piezoelectric tensor is
\begin{equation}
c_{ijk}^{(b)}={dP_i^{(b)}\over d\epsilon_{jk}}
\;\;.
\label{eq:improperb}
\end{equation}
Since $\bf P$ is well-defined modulo $e{\bf R}/\Omega$, and
both $\bf R$ and $\Omega$ vary with the deformation $\epsilon$,
Eq.~(\ref{eq:improperb}) will clearly give different results
for different choices of branch.  This branch-dependence
is problematic; the piezoelectric tensor is measurable, and a
suitable theory ought to give a unique value for it.

Before proceeding, the reader is reminded that the piezoelectric
response contains, in general, a ``clamped-ion'' part and an
``internal-strain'' part \cite{sck,bernardini,waghmare}.
That is, one decomposes the actual deformation into a sum
of two parts: a homogeneous strain in which the nuclear
coordinates follow Eq.~(\ref{eq:deform}) exactly (clamped-ion part),
plus an internal distortion of the nuclear coordinates at fixed
strain (internal-strain part).  Since the latter occurs at fixed
strain, all the subtleties about the branch-dependence and the
proper-vs.-improper distinction disappear for this case.  While
the computation of the internal-strain part of the piezoelectric
response may be tedious (requiring an iterative set of
force calculations to determine the needed internal relaxations),
it is straightforward in principle.  Consequently, for the
remainder of this paper, the discussion refers to the clamped-ion
response only unless explicitly stated otherwise.


\subsection{Branch-invariance of proper piezoelectric response}
\label{sec:branch}

While it is true that the improper piezoelectric response depends,
in general, on choice of branch, it is instead the {\it proper}
piezoelectric tensor that should be compared with experiment.
Figure 2 shows a sketch of one possible experimental setup, in which
a block of piezoelectric material is sandwiched between shorted
conducting electrodes, and the current $I$ that flows in response
to a deformation $\epsilon$ is measured.  As suggested by
Eq.~(\ref{eq:proper}), the proper piezoelectric response is
related to the current that flows through the sample in response
to the deformation, and is thus the experimentally
measured quantity.  Moreover, the induced current density
${\bf j}(\bf r)$ is periodic with the lattice, so that its unit
cell average $\bf J$ in Eq.~(\ref{eq:proper}) is perfectly
well-defined, and consequently the proper piezoelectric tensor
$\widetilde{c}$ cannot suffer from any dependence upon choice
of branch.

\begin{figure}
\epsfxsize=2.0in
\centerline{\epsfbox{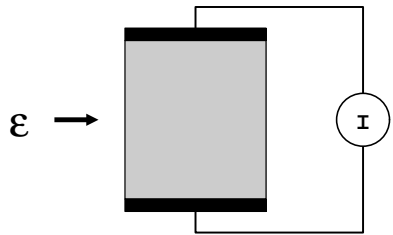}}
\vskip 0.5cm
\caption{Sketch of experiment for measuring proper piezoelectric
coefficients.  Strain $\epsilon$ is applied to piezoelectric
material (gray) sandwiched between grounded capacitor plates, and
resulting current $I$ is measured.}
\label{fig2}
\end{figure}

It is straightforward to check this branch-independence of
$\widetilde{c}$ explicitly.  Since the polarizations for two
different branch choices are related by
\begin{equation}
{\bf P}^{(b')}={\bf P}^{(b)}+{e\over\Omega}{\bf R}
\;\;,
\label{eq:pp}
\end{equation}
one finds
\begin{eqnarray}
dP_i^{(b')}&=&dP_i^{(b)}\,-\,{e\over\Omega^2}\,d\Omega\,R_i \,+\,
   {e\over\Omega}\,dR_i \nonumber\\
&=& dP_i^{(b)}\,+\,{e\over\Omega}\,\sum_l\,
   (-d\epsilon_{ll}\,R_i + d\epsilon_{il}\,R_l)
\;\;.
\label{eq:dP}
\end{eqnarray}
so that
\begin{equation}
c_{ijk}^{(b')}=c_{ijk}^{(b)} - {e\over\Omega}\delta_{jk}R_i
  + {e\over\Omega}\delta_{ij}R_k
\;\;,
\label{eq:cbp}
\end{equation}
or, using Eq.~(\ref{eq:pp}),
\begin{equation}
c_{ijk}^{(b')}+\delta_{jk}P^{(b')}_i-\delta_{ij}P^{(b')}_k =
c_{ijk}^{(b)} +\delta_{jk}P^{(b)}_i +\delta_{ij}P^{(b)}_k
\;\;.
\label{eq:proof}
\end{equation}
It is thus evident that $\widetilde{c}_{ijk}$
as defined in Eq.~(\ref{eq:relation}) is indeed independent of
choice of branch.

It is instructive to note that a similar argument applies to the part
of the proper piezoelectric tensor arising from the ionic contribution
${\bf P}_{\rm ion}=(e/\Omega)\sum_\tau Z_\tau {\bf r}_\tau$ in
Eq.~(\ref{eq:Ptot}).  Recalling that we are working in the clamped-ion
approximation, so that $d{\bf r}_\tau$ follows the form of
Eq.~(\ref{eq:deform}), one finds immediately that
$\widetilde{c}_{\rm ion}=0$ by the same logic as for the previous
paragraph.

Indeed, the same logic would apply to Eq.~(\ref{eq:Ptotnew}) if
the Wannier centers ${\bf r}_n$ would undergo a homogeneous
deformation of the type (\ref{eq:deform}).  In other words,
{\it the proper piezoelectric response is identically zero for
a homogeneous deformation of both the ionic positions and the
Wannier centers}, in which case there is no charge flow through the
interior of the crystal.


\subsection{Simplified finite-difference formula}
\label{sec:formula}

Of course, there is no reason to expect the Wannier centers ${\bf
r}_n$ to follow a homogeneous deformation, so $\widetilde{c}$ is
not generally zero.  But from this point of view, it becomes
evident that the {\it the proper piezoelectric response is precisely
a measure of the degree to which the Wannier centers fail to follow a
homogeneous deformation.}  Or equivalently, returning to
Eq.~(\ref{eq:buildP}), we see that the proper piezoelectric
response measures just the variation of the Berry phases
$\phi_{n,\alpha}$ with the strain deformation.  More precisely,
starting from Eqs.(\ref{eq:buildP}), (\ref{eq:improper}), and
(\ref{eq:relation}), one finds
\begin{equation}
\widetilde{c}_{ijk} \,=\,-{1\over2\pi}\,{e\over\Omega}
   \sum_{n,\alpha} \,{d\phi_{n,\alpha}\over d\epsilon_{jk}} \,
   R_{\alpha i}
\label{eq:simple}
\end{equation}
We have been working in the clamped-ion approximation, but in
general, if there are internal relaxations accompanying the
deformation, one can define a total Berry phase in direction
$\alpha$,
\begin{equation}
\phi_\alpha \,=\, \sum_\tau \, Z_\tau\,{\bf G}_\alpha\cdot{\bf r}_\tau
   \,-\,\sum_n \, \phi_{n,\alpha}
\;\;,
\label{eq:phitot}
\end{equation}
so that
\begin{equation}
\widetilde{c}_{ijk} \,=\,{1\over2\pi}\,{e\over\Omega}
   \sum_{\alpha} \,{d\phi_\alpha\over d\epsilon_{jk}} \,
   R_{\alpha i}
\;\;.
\label{eq:isimple}
\end{equation}
Naturally, the ionic contributions to $d\phi_\alpha/d\epsilon_{jk}$
vanish in the clamped-ion approximation.

Equation (\ref{eq:simple}), or its generalization (\ref{eq:isimple}),
is the central result of this paper, and provides a simple and
practical recipe for calculating the desired proper piezoelectric
response.  One simply computes the needed $d\phi/d\epsilon$ by
finite differences, as $(\phi'-\phi)/(\epsilon'-\epsilon)$ for
nearby strain configurations $\epsilon$ and $\epsilon'$.  Then
these $d\phi/d\epsilon$ are inserted into Eq.~(\ref{eq:simple})
or (\ref{eq:isimple}) to obtain the elements of the proper piezoelectric
tensor.


\subsection{Relation to surface charges}
\label{sec:surface}

At the end of Sec.~\ref{sec:branch}, it was pointed out that a
homogeneous deformation of the lattice of positive ionic
and negative Wannier-center point charges would give rise to no
internal current, and hence no proper piezoelectric response.
This result can be made more intuitive by considering the
connection between bulk polarization and surface charges \cite{vks}.

\begin{figure}
\epsfxsize=2.5in
\centerline{\epsfbox{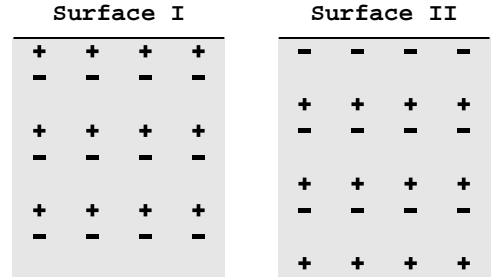}}
\vskip 0.5cm
\caption{Two possible surface terminations of the lattice of point
charges shown in Fig.~1(b).}
\label{fig3}
\end{figure}

Consider, for example, a crystallite composed of $N\times N\times N$
replicas of the unit cell shown in Fig.~1(b).  In general there
may be an arbitrariness in the choice of surface termination, as
illustrated for the top surface of this crystallite in Fig.~3.
For any given termination, the macroscopic surface charge density
$\sigma$ is uniquely defined as $\int dz\,\overline{\rho}(z)$,
where $\overline{\rho}(z)$ is the average charge contained in a
unit cell centered at vertical coordinate $z$ (so that
$\overline{\rho}$ vanishes either deep in the crystal or deep in
the vacuum and its integral is convergent).
For the crystal of Fig.~1, one finds $\sigma=\gamma
e/a^2$ and $\sigma=(\gamma-1)e/a^2$ for the terminations of type
I and II of Figs.~3(a) and 3(b), respectively.  Referring
back to Sec.~\ref{sec:berry}, where it was found that
$P_z=(\gamma+m_3)e/a^2$, one confirms that the relation \cite{vks}
\begin{equation}
\sigma={\bf P}\cdot\hat{\bf n}
\label{eq:surf}
\end{equation}
is satisfied for both terminations, the ambiguity of termination
corresponding to the choice of branch of $\bf P$.

For definiteness, assume that the surface terminations are such that
the top and bottom surfaces of the crystallite have charge
densities $+\gamma e/a^2$ and $-\gamma e/a^2$ on the top and bottom
surfaces, respectively, and zero on the sides.  Then the magnitude
of the total charge on the top or bottom surface is just
$N^2a^2\sigma=N^2\gamma e$, which is clearly independent of any
homogeneous ($\gamma$-preserving) deformation of the crystal.
Thus, if this crystallite were inserted between grounded
capacitor plates as in Fig.~2, {\it no current would flow}
through the wire as a result of the homogeneous deformation.  This
is consistent with the vanishing of the proper piezoelectric
response associated with such a homogeneous deformation, as
already illustrated via Eq.~(\ref{eq:simple}).

However, for the same situation, the {\it improper} piezoelectric
tensor {\it would} have nonzero elements.  For the chosen surface
termination, the crystallite has a total dipole moment
${\bf d}=N^3\gamma ec\hat{\bf z}$, and a polarization
${\bf P}={\bf d}/N^3a^2c=(\gamma e/a^2)\hat{\bf z}$ as
expected.  Clearly this $\bf P$ is invariant with respect to
an elongation of the crystallite along the $\hat{\bf z}$ axis
(strain component $\epsilon_{zz}$), but {\it not} to an
elongation along the $\hat{\bf x}$ or $\hat{\bf y}$ axes
($\epsilon_{yy}$ or $\epsilon_{zz}$), thus explaining why
there is a correction to $c_{zxx}$ but not to $c_{zzz}$ in
Eq.~(\ref{eq:writeout}).  Similar considerations applied to
shear strains and rotations explain the remaining entries in
Eq.~(\ref{eq:writeout}).


\section{Discussion}
\label{sec:disc}

As is evident from Eqs.(\ref{eq:relation}) and (\ref{eq:writeout}),
the distinction between the proper and improper piezoelectric tensor
is only present if a spontaneous polarization is present.  If the
spontaneous polarization is small, as for wurtzite semiconductors
\cite{bernardini}, it may be a good approximation to neglect
the corrections to the improper tensor components.  Alternatively,
linear-response methods can be used to compute the proper piezoelectric
response directly \cite{waghmare}.  However, for a finite-difference
calculation of the proper piezoelectric response of a ferroelectric
material, it is essential to take the corrections to the improper
response explicitly into account, as was done in Ref.~\cite{sck}.


\section{Summary}
\label{sec:summary}

In this work, a simple and straightforward method for computing
the proper piezoelectric response has been proposed.  Instead of
first computing the improper response and then the needed corrections,
the proper response is computed directly from Eq.~(\ref{eq:simple})
or (\ref{eq:isimple}).  It is thus clarified that the central
quantities needed to determine the proper piezoelectric
response are just the variations of the Berry phases with
deformation.


\acknowledgments

This work was supported by ONR Grant N0001497-1-0048.
I wish to thank R.~Cohen for calling my attention to the problem
of the proper piezoelectric response and its relation to the
Berry-phase theory of polarization.


\end{document}